\documentstyle[12pt,preprint]{aastex}  

\begin{document}

\title{A Southern Sky and Galactic Plane Survey for Bright Kuiper Belt Objects}
\author{Scott S. Sheppard\altaffilmark{1}, Andrzej Udalski\altaffilmark{2}, Chadwick Trujillo\altaffilmark{3}, Marcin Kubiak\altaffilmark{2}, Grzegorz Pietrzynski\altaffilmark{2}, Radoslaw Poleski\altaffilmark{2}, Igor Soszynski\altaffilmark{2}, Michal K. Szyma{\'n}ski\altaffilmark{2}, and Krzysztof Ulaczyk\altaffilmark{2}}    

\altaffiltext{1}{Department of Terrestrial Magnetism, Carnegie Institution of Washington, 5241 Broad Branch Rd. NW, Washington, DC 20015, USA, sheppard@dtm.ciw.edu}
\altaffiltext{2}{Warsaw University Observatory, Al. Ujazdowskie 4, 00-478 Warszawa, Poland}
\altaffiltext{3}{Gemini Observatory, 670 North A'ohoku Place, Hilo, HI 96720, USA}


\begin{abstract}  

About 2500 square degrees of sky south of declination -25 degrees
and/or near the galactic plane were surveyed for bright outer solar
system objects.  This survey is one of the first large scale southern
sky and galactic plane surveys to detect dwarf planets and other
bright Kuiper Belt objects in the trans-Neptunian region.  The survey
was able to obtain a limiting R-band magnitude of 21.6.  In all, 18
outer solar system objects were detected, including Pluto which was
detected near the galactic center using optimal image subtraction
techniques to remove the high stellar density background.  Fourteen of
the detections were previously unknown trans-Neptunian objects,
demonstrating that the southern sky had not been well-searched to date
for bright outer solar system objects.  Assuming moderate albedos,
several of the new discoveries from this survey could be in
hydrostatic equilibrium and thus be considered dwarf planets.
Combining this survey with previous surveys from the northern
hemisphere suggests that the Kuiper Belt is nearly complete to around
21st magnitude in the R-band.  All the main dynamical classes in the
Kuiper Belt are occupied by at least one dwarf planet sized object.
The 3:2 Neptune resonance, which is the innermost well-populated
Neptune resonance, has several large objects while the main outer
Neptune resonances such as the 5:3, 7:4, 2:1, and 5:2 do not appear
have any large objects.  This indicates that the outer resonances are
either significantly depleted in objects relative to the 3:2 resonance
or have a significantly different assortment of objects than the 3:2
resonance.  For the largest objects ($H<4.5$ mag), the scattered disk
population appears to have a few times more objects than the main
Kuiper Belt population, while the Sedna population could be several
times more than that of the main Kuiper Belt.

\end{abstract}

\keywords{Kuiper Belt -- Oort Cloud -- comets: general -- minor planets, asteroids -- solar system: general -- planetary formation}

\section{Introduction}

The strong dynamical connection that the trans-Neptunian objects
(TNOs) have to the planets makes determining their population and
orbital structures valuable for gaining insight into solar system
formation and planet evolution.  The Kuiper Belt, a remnant of the
original protoplanetary disk, has a ``fossilized'' record of the
original solar nebula and subsequent evolution of the solar system.
TNOs are likely primitive with significant amounts of volatiles.  The
largest TNOs or dwarf planet sized objects are rare but extremely
important for several reasons: 1) The brightest few objects are the
only ones accessible to high signal to noise spectroscopy techniques
that are required to determine surface compositions, such as methane
and water ice (Barucci et al. 2008; Trujillo et al. 2011).  These
physical characteristics are important in order to understand the
formation, origin and composition of the objects and gain insight into
planet formation and chemistry in the original solar nebula. 2) The
size distribution of the biggest objects in the Kuiper Belt determines
if the mass in the Kuiper Belt is dominated by the largest or smallest
objects, which is a key metric of planetismal growth scenarios (Kenyon
et al. 2008, 2010; Cuzzi et al. 2010).  The size and number of the
biggest objects constrain the density and thus planet formation
ability of the original solar nebula in the outer solar system.  3)
Occultations of stars by the biggest TNOs are possible to predict and
observe in order to probe the TNOs sizes, shapes, albedos, and
atmospheres (Elliot and Kern 2003; Elliot et al. 2010).

The Palomar 48 inch Schmidt telescope in the northern hemisphere, with
one of the largest CCD cameras in the world, was used to survey most
of the sky north of -25 degrees declination for the brightest ($m_{R}
\lesssim 21$ mags) TNOs (Trujillo and Brown 2003; Brown et al. 2004,
2005; Brown 2008; Schwamb et al. 2009, 2010).  In these surveys tens
of bright TNOs including likely dwarf planets Eris, Makemake, Haumea,
Orcus, Quaoar, Sedna and 2007 OR10 were discovered.  These surveys
showed that many of the largest Kuiper Belt Objects (KBOs) have
relatively large inclinations with the vast majority of KBOs expected
to be found within about 20 degrees of the ecliptic (Brown 2008).
Extrapolating the Cumulative Luminosity Function (CLF) to the bright
end of the KBOs indicates several large KBOs should be discovered in
the southernmost parts of the sky that the surveys from the northern
telescopes did not image.

The southern hemisphere has not been well-surveyed for distant solar
system objects until now because in the past there were no sensitive,
wide-field digital imagers on suitable telescopes in the south.  This
changed in 2009 when a large wide-field imager was put onto the 1.3
meter Warsaw telescope at Las Campanas in Chile.  The OGLE-Carnegie
Kuiper Belt Survey (OCKS) was implemented to search the Kuiper Belt
for dwarf planets and bright TNOs through a shallow survey to fainter
than 21st magnitude in the R-band from the southern hemisphere.  OCKS
covered the area within a few tens of degrees of the ecliptic for
declinations less than -25 degrees and the crowded galactic plane
fields in the north and south.  Another independent southern sky
survey for KBOs was started in late 2009 with the Schmidt telescope at
La Silla (Rabinowitz 2010).  This is the first time most of this sky
area was searched for outer solar system objects with modern digital
CCD detectors.

\section{Observations}

The vast majority of the survey fields were obtained with the Warsaw
1.3 meter telescope at Las Campanas observatory in Chile.  The
telescope is also known as the OGLE telescope (Optical Gravitational
Lensing Experiment; Udalski et al. 1994) and OCKS is considered part
of the OGLE-IV project.  OGLE-IV commenced with the successful
commissioning of the new wide-field 1.4 square degree imager at the
beginning of 2010. The southern sky Kuiper Belt survey observations at
the Warsaw telescope occurred between March and September 2010 while
the northern galactic plane fields near the ecliptic were imaged in
December 2010 and January 2011. The 1.4 square degree imager has 32
E2V44-82 $2048\times 4102$ CCD chips with $0.\arcsec 26$/pixel.  There
are four rows and 9 columns of chips.  Gaps are generally only a few
arcseconds between chips except between the first and second rows and
third and fourth rows the gaps are a bit wider at several tens of
arcseconds.  Readout time for the detector is about 20 seconds.

All fields were within about 2.5 hours of opposition with most being
within 1.5 hours.  At these opposition distances, the apparent motion
of an outer solar system object is dominated by the parallax from the
Earth's movement, making confusion of outer solar system objects with
foreground main belt asteroids minimal (Luu and Jewitt 1988).  Las
Campanas is a very dark site with excellent seeing conditions
(Thomas-Osip et al. 2011).  Most images were obtained with the seeing
around 1 arcsecond or less.  If the seeing was much worse than 1
arcsecond or if the conditions were not photometric on a given night,
observations were not taken.  Integrations were 180 seconds with the
telescope tracking at sidereal rates.  Since there was no preferred VR
or R-band filter for the 1.4 square degree imager, a V-band filter was
used at the start of the survey for fields West of the Galactic plane.
Because of the better seeing conditions in the I-band, the I-band
filter was used for fields in the Galactic plane as well as fields
East of the Galactic plane.  It was found that the V-band and I-band
images obtained similar depths but the I-band was preferred since it
was less sensitive to moderate moon brightness.  Image reduction was
performed by first bias subtracting and then flat-fielding the images.

In addition to the Warsaw data, about 100 square degrees were surveyed
using the CTIO 4 meter Blanco telescope with its MOSAIC II camera that
covers about a third of a square degree.  These data were obtained in
June 2009 and 2010 in order to see how well such a program would work
on the 4 meter telescope.  Images were only 20 seconds in length and
reached magnitudes of about 22 in the R-band.  Recovery was mostly
done at the Warsaw 1.3 meter telescope but some recovery took place at
the CTIO 4 meter and Magellan 6.5 meter.

\section{Analysis}

\begin{figure}
\epsscale{0.4}
\centerline{\includegraphics[angle=0,totalheight=0.6\textheight]{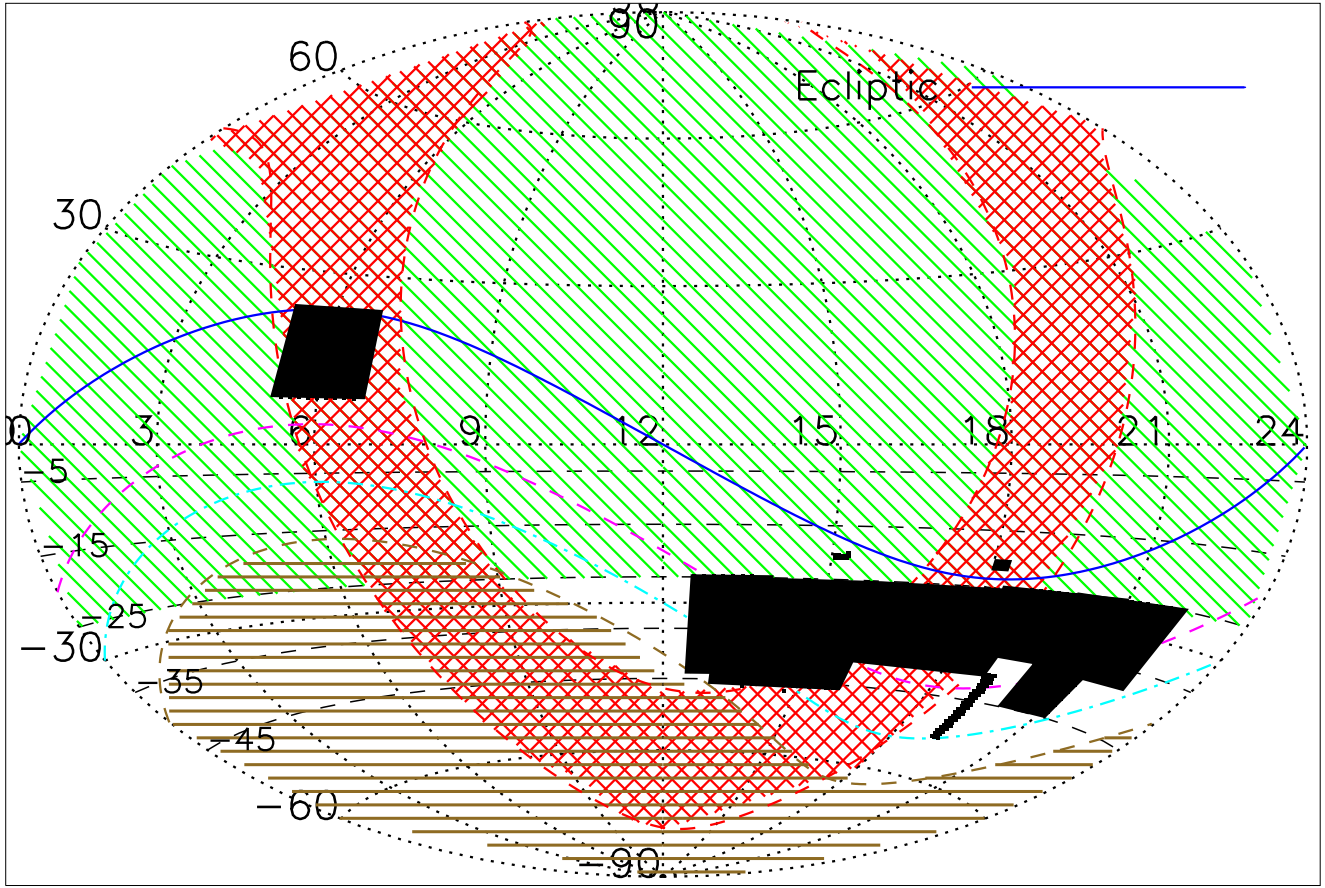}}
\caption{The black shaded regions represent the sky area surveyed in
  this work.  The horizontal axis is the right ascension in hours and
  the vertical axis is the declination in degrees.  The blue solid
  line shows the ecliptic, the purple dashed line shows $-20$ degrees
  from the ecliptic and the cyan dotted dashed line shows $-30$
  degrees from the ecliptic.  Areas more than $-40$ degrees south of
  the ecliptic are shown with brown horizontal stripes.  The area
  within 15 degrees of the galactic plane is shown with red crossed
  stripes while the area covered by wide-field KBO surveys from the
  north (Trujillo and Brown 2003; Brown 2008; Schwamb et al. 2009,
  2010) are shown with green angled stripes.  Almost all KBOs are
  expected to be within 20 degrees of the ecliptic with it highly
  unlikely any KBO is beyond 40 degrees from the ecliptic (Brown
  2008).}
\label{fig:MapOgle} 
\end{figure}

In total about 2500 square degrees of sky were surveyed in the
southern hemisphere or near the galactic plane
(Figure~\ref{fig:MapOgle}).  Each survey field had at least two hours
between the first and last image of a three image sequence.  Outer
solar system objects were searched for in the survey fields in two
complementary ways.  One technique used a computer algorithm
specifically designed to detect the apparent motion of trans-Neptunian
objects (Trujillo and Jewitt 1998; Sheppard and Trujillo 2010) while a
second technique used a differencing algorithm (Udalski et
al. 1997,2003; Wozniak 2000) on the three images in order to remove
the steady state of background stars to look for moving or transient
objects.  The differencing algorithm was used on all fields and was
the only technique used on fields within 15 degrees of the galactic
plane.

Both computer algorithms were calibrated to detect moving objects that
appeared in all three images from one night and had a motion
consistent with being beyond 10 AU (motion slower than about 10
arcseconds per hour).  Because of the fine pixel scale and relatively
good seeing, the survey was sensitive to objects moving as slow as
$0.5$ arcseconds per hour.  This apparent motion corresponds to
objects out to about 300 AU.  Since the survey covers many nights, the
data from night to night is not all of the same quality.  In order to
make the data as consistent as possible over the nights, survey fields
were only taken in moderate seeing ($\sim 1$ arcsecond) or better
conditions and only when conditions were photometric.  If the seeing
was significantly worse than about one arcsecond, the survey was not
continued for that night.

\begin{figure}
\epsscale{0.4}
\centerline{\includegraphics[angle=90,totalheight=0.6\textheight]{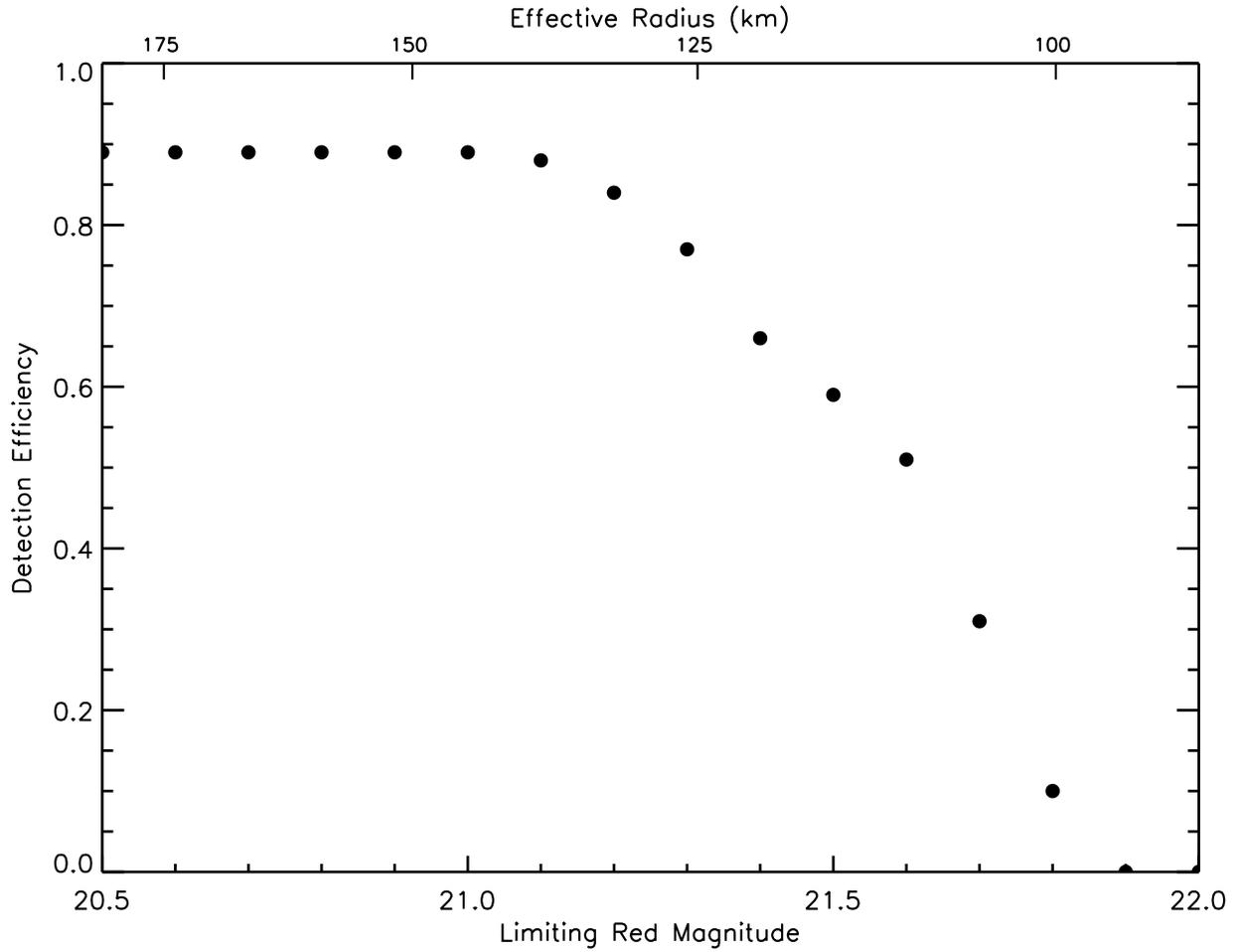}}
\caption{Detection efficiency of the KBO survey versus the apparent
  red magnitude using the Warsaw 1.3 meter telescope.  In good seeing
  (0.8 arcseconds FWHM) the $50\%$ detection efficiency is at about
  21.6 mags while in moderate seeing ($\sim 1$ arcsecond) it is about
  21.2 mags in the R-band.  Effective radii of the apparent magnitudes
  were calculated assuming the object has an albedo of 0.15 and is at
  40 AU.}
\label{fig:limitingmag} 
\end{figure}

The limiting magnitude of the survey was determined by placing
artificial objects in the fields matched to the point spread function
of the images with motions mimicking that of a TNO ($4$ to $0.5$
arcseconds per hour).  A $50 \%$ detection efficiency at an R-band
limiting magnitude of about 21.6 magnitudes was found for fields with
good seeing conditions about 15 degrees or more from the galactic
plane (Figure \ref{fig:limitingmag}).  For fields with moderate seeing
conditions the R-band limiting magnitude was found to be about 21.2
magnitudes, where the typical color of a moderately red KBO was used
to convert the I-band survey fields to the R-band (R-I=0.5 mags) in
order to better compare the survey with previous survey depths.

\begin{figure}
\epsscale{0.4}
\centerline{\includegraphics[angle=0,totalheight=0.6\textheight]{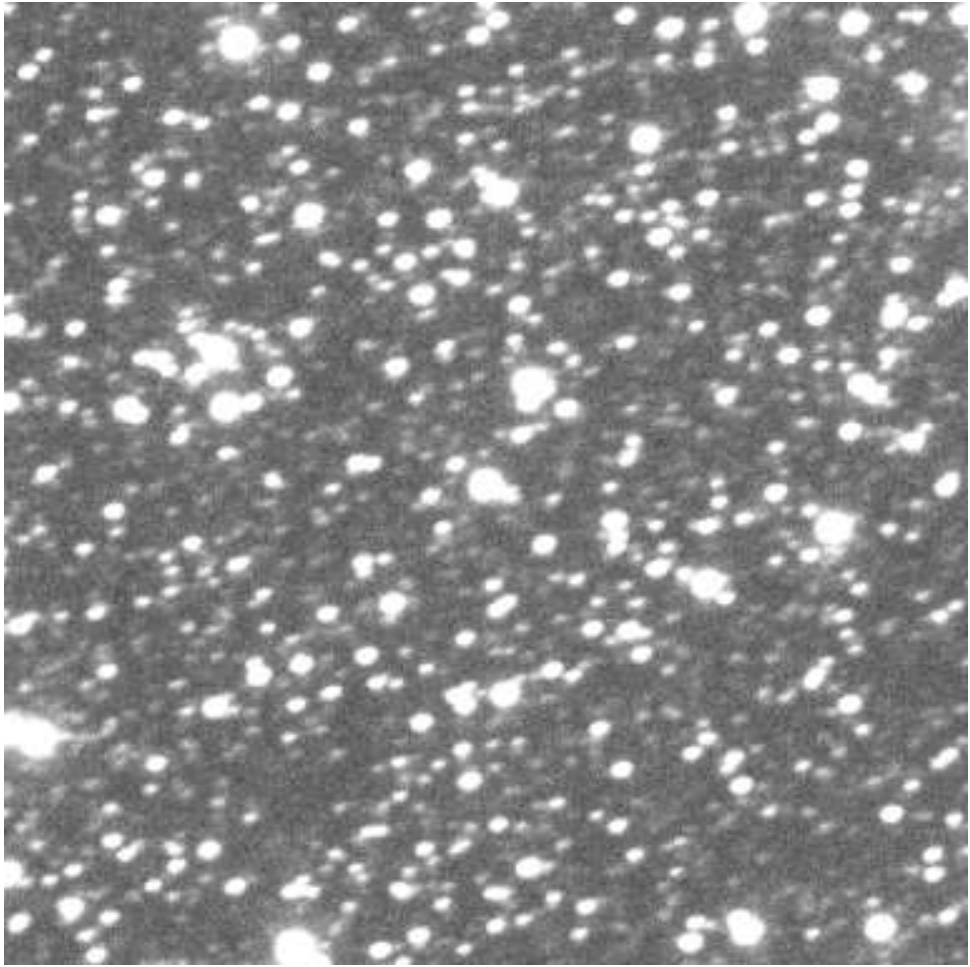}}
\caption{A small portion of an image showing Pluto in the galactic
  plane from the 1.3 meter Warsaw telescope.  Pluto is in the center
  of this image as revealed in Figure~\ref{fig:plutodiff}.}
\label{fig:pluto1} 
\end{figure}

\begin{figure}
\epsscale{0.4}
\centerline{\includegraphics[angle=0,totalheight=0.6\textheight]{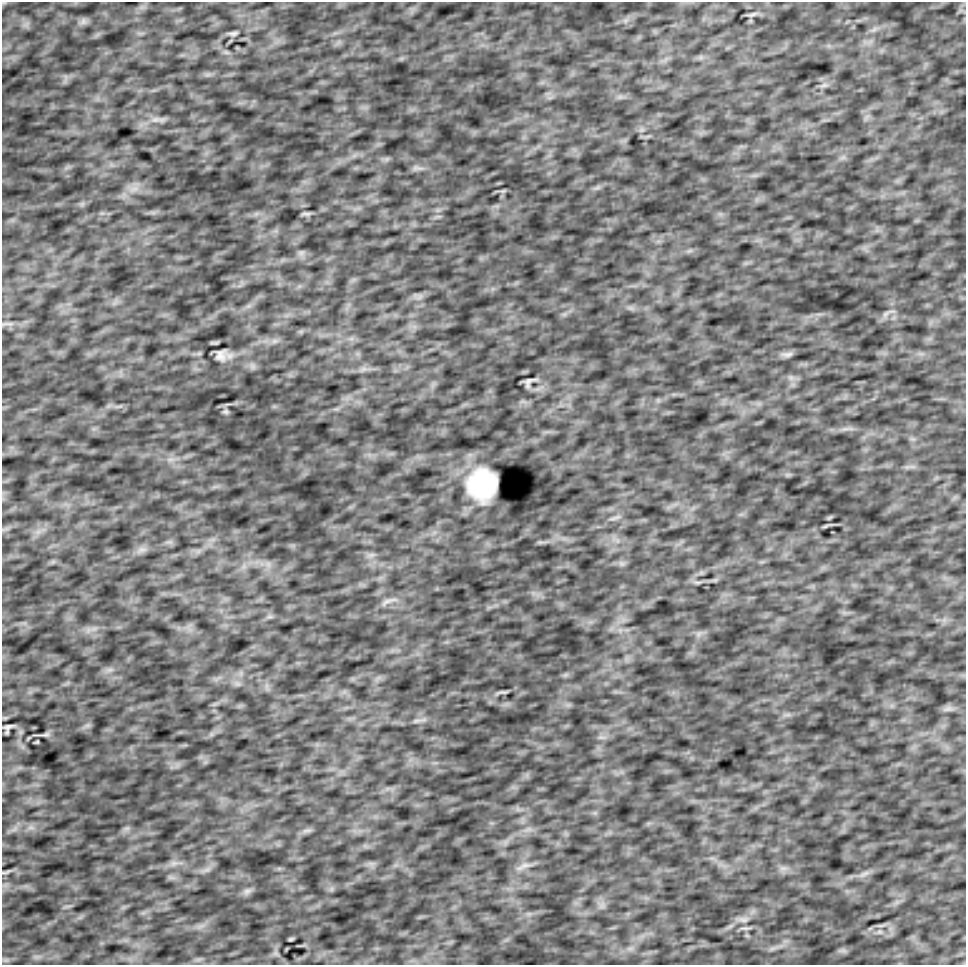}}
\caption{A difference image of the Pluto fields
  (Figure~\ref{fig:pluto1}) showing the removal of the steady state
  background of stars.  The motion of Pluto is clearly revealed in the
  difference image as a positive (bright) and negative (dark) point
  from the subtraction process of the two individual images.}
\label{fig:plutodiff} 
\end{figure}

For images near the galactic plane the stellar confusion would limit
the detection of moving solar system objects in previous surveys.  In
this survey the optimal PSF matching image subtraction techniques
developed by Alard and Lupton (1998) and Alard (2000) and implemented
through the previous OGLE phases were used (Wozniak 2000; Wozniak et
al. 2001; Udalski 2003).  PSF matching and image subtraction removed
the stellar confusion from the galactic plane.  Thus, this is the
first survey to be sensitive to TNOs near the galactic center where
the ecliptic plane crosses the galaxy.  To test the moving object
algoritm with differenced images, Pluto was observed early in the
survey and easily found in the dense galactic plane
(Figures~\ref{fig:pluto1} and~\ref{fig:plutodiff}).  The survey depth
near the galactic center was similar to the depth of the fields off
the galactic plane, but the survey efficiency of detection was
decreased by about 15 percent.

\section{Results and Discussion}

\subsection{Completion Limits of the Kuiper Belt}

Eighteen outer solar system objects were detected in this survey.
Fourteen of these objects were new discoveries showing that this
region of sky had not been well-searched for bright, distant objects
in the past (Tables 1 and 2).  Combining this southern sky and
galactic plane survey with the previous large area northern sky
surveys (Trujillo and Brown 2003; Brown 2008; Schwamb et al. 2009,
2010) and a recent large Kuiper Belt survey in the south started by
Rabinowitz (2010) makes it likely that the Kuiper Belt is now nearly
complete to about 21st magnitude in the R-band.  To date, only three
areas have not been well searched for bright outer solar system
objects: 1) southern fields very distant from the ecliptic ($>20$
degrees ecliptic latitude) and thus unlikely to harbor many bright
KBOs, 2) a few hundred square degrees in the northern section of the
north galactic plane near the ecliptic and 3) a few hundred square
degrees in the northern section of the south galactic plane near the
ecliptic (Figure~\ref{fig:MapOgle}).  There are around 70 known TNOs
with apparent magnitudes brighter than 21 in the R-band (see section
4.2), almost all within 20 degrees of the ecliptic. Thus, there is on
average one KBO brighter than 21st mag every few hundred square
degrees of sky near the ecliptic.  This means there is likely to be
only one or two KBOs brighter than 21st magnitude in the few remaining
areas yet to be searched.  Though nearly complete to 21st magnitude
now, some objects, especially Centaurs and scattered disk objects,
have large eccentricities and thus could become brighter than 21st
magnitude in the future as they approach perihelion.

\begin{figure}
\epsscale{0.4}
\centerline{\includegraphics[angle=0,totalheight=0.6\textheight]{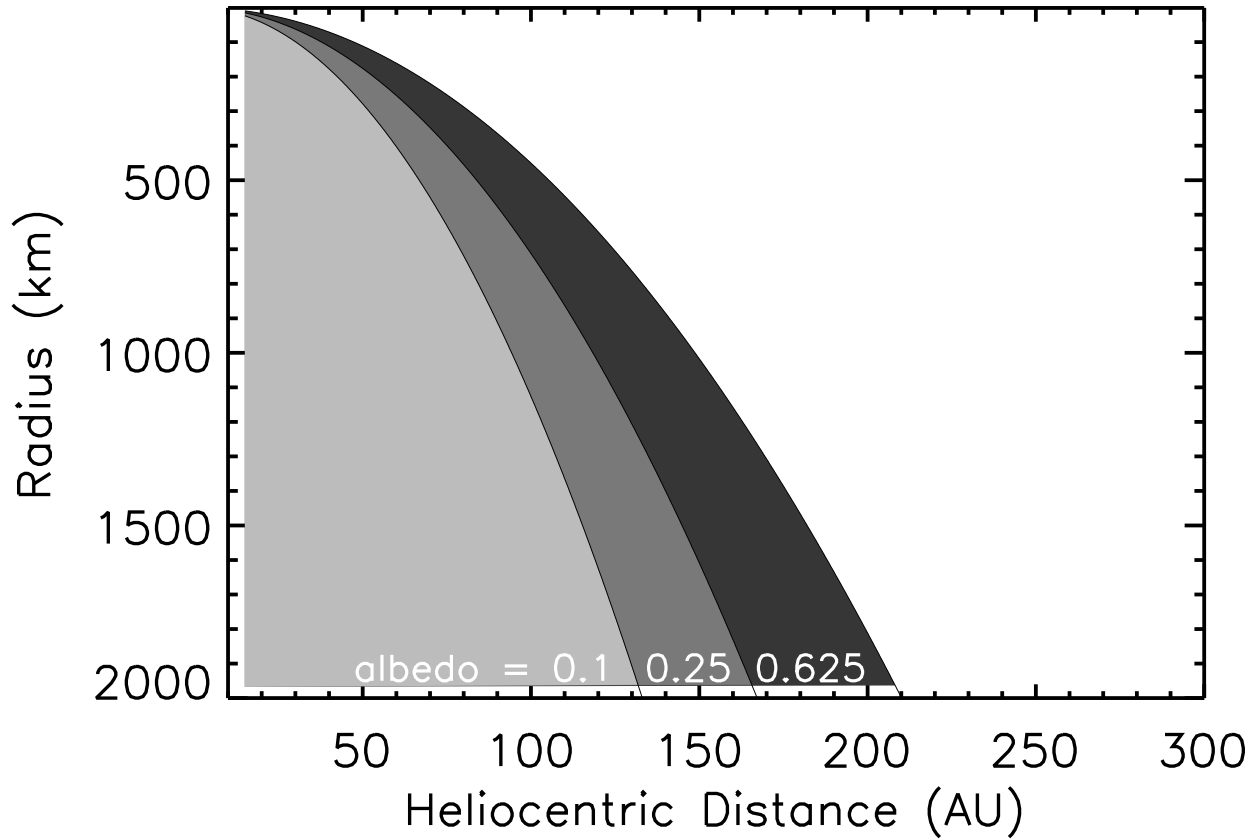}}
\caption{The outer solar system has now been surveyed to a
  completeness limit of about 21 magnitudes in the R-band.  This
  figure shows what size and distance an object would be for several
  different albedos for a 21st magnitude object.  Shaded areas
  correspond to completeness limits for albedo $<0.1$ (light), albedo
  $<0.25$ (medium) and albedo $<0.625$ (dark).  It is clear that a
  Pluto (1161 km) or even larger sized object could easily have gone
  undetected to date if beyond a few hundred.}
\label{fig:distancesize21} 
\end{figure}

\begin{figure}
\epsscale{0.4}
\centerline{\includegraphics[angle=0,totalheight=0.6\textheight]{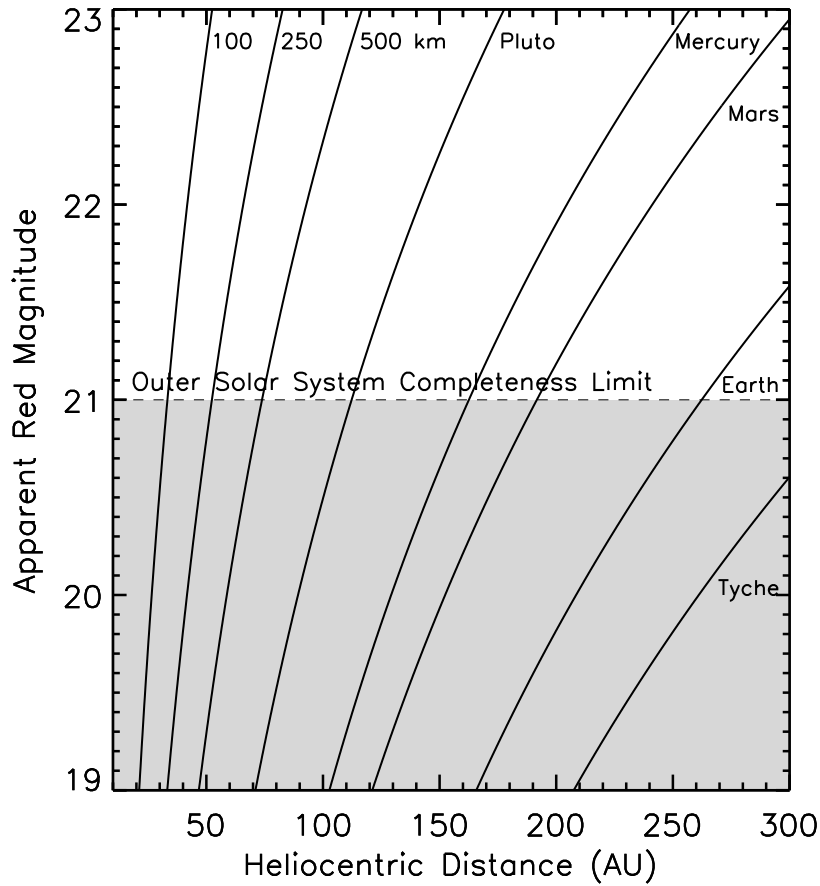}}
\caption{The radius of an object is shown assuming a moderate albedo
  of 0.15 for various heliocentric distances and apparent red
  magnitudes.  The known objects in the Kuiper Belt region are
  complete to about 21st magnitude, shown by the shaded region below
  the dashed line.  It is likely that everything under the dashed line
  at 21 magnitudes is known.  The radii used for the named objects in
  the figure are Pluto (1161 km), Mercury (2440 km), Mars (3396 km),
  Earth (6371 km) and an arbitrary lower limit on a hypothetical
  eccentric giant planet or companion to our Sun, sometimes called
  Nemesis or Tyche, with 10,000 km radius (Iorio 2009; Melott and
  Bambach 2010).  In the distant solar system very large objects would
  easily be undetected to date.}
\label{fig:distanceogle} 
\end{figure}

\begin{figure}
\epsscale{0.4}
\centerline{\includegraphics[angle=0,totalheight=0.6\textheight]{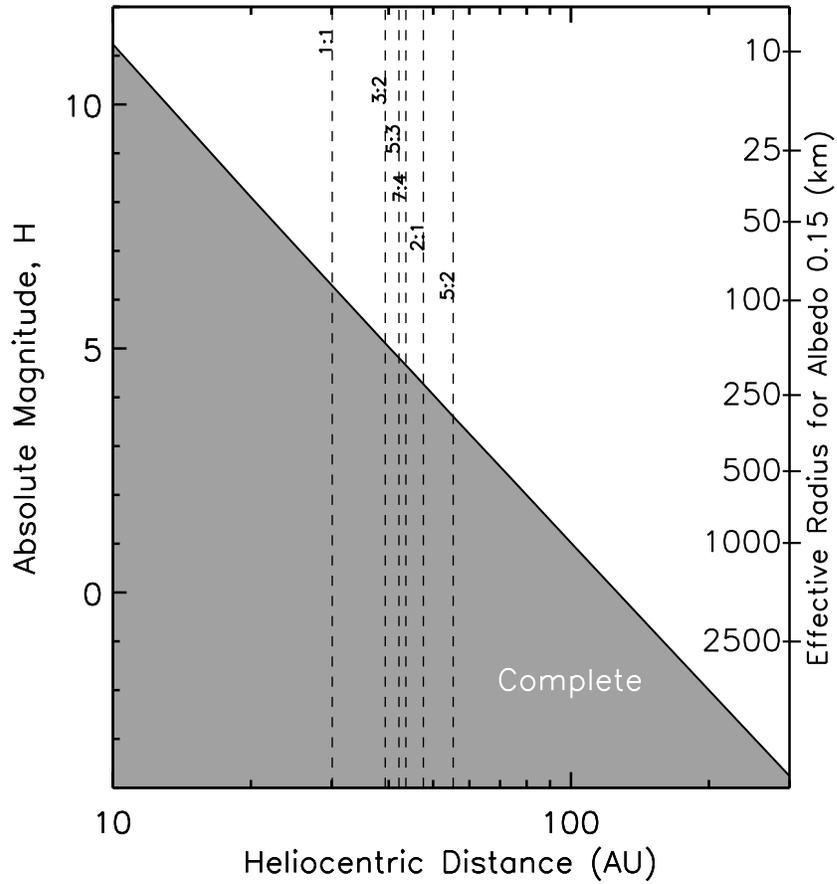}}
\caption{The heliocentric distance versus the completeness of absolute
  magnitude, H.  The shaded region shows where the outer solar system
  should be complete in discoveries.  The effective radius on the
  right side assumes an albedo of 0.15.  The average semi-major axis
  for the various major Neptune resonance populations are shown as
  vertical dashed lines for reference.}
\label{fig:distanceH} 
\end{figure}

The size of an object at the completeness limit depends on the
distance and albedo of the object (Figure~\ref{fig:distancesize21}).
The largest few objects, with radii greater than about 500 km, have
been found to have very high albedos ($\rho_{R} \sim 0.6-0.8$), while
smaller objects appear to have moderate albedos ($\rho_{R} \sim
0.1-0.2$) (Stansberry et al. 2008).  The high albedos of the largest
objects is likely due to atmospheres and/or surface processes such as
cryovolcanism (Licandro et al. 2006; Dumas et al. 2007; Rabinowitz et
al. 2007; Sheppard 2007).  Assuming a typical albedo of
$\rho_{R}=0.15$ for the moderate sized KBOs of 21st magnitude, the
completeness limit at 30 AU is about 80 km in radius, while at 50 AU
it is about 225 km in radius (Figure~\ref{fig:distanceogle}).  In
absolute magnitude, H, this would be 6.6 and 4.4 magnitudes
respectively (Figure~\ref{fig:distanceH}).  It is clear that further
Pluto or larger sized objects could remain undetected if beyond a few
hundred AU.

\subsection{Size Distribution}

\begin{figure}
\epsscale{0.4}
\centerline{\includegraphics[angle=0,totalheight=0.6\textheight]{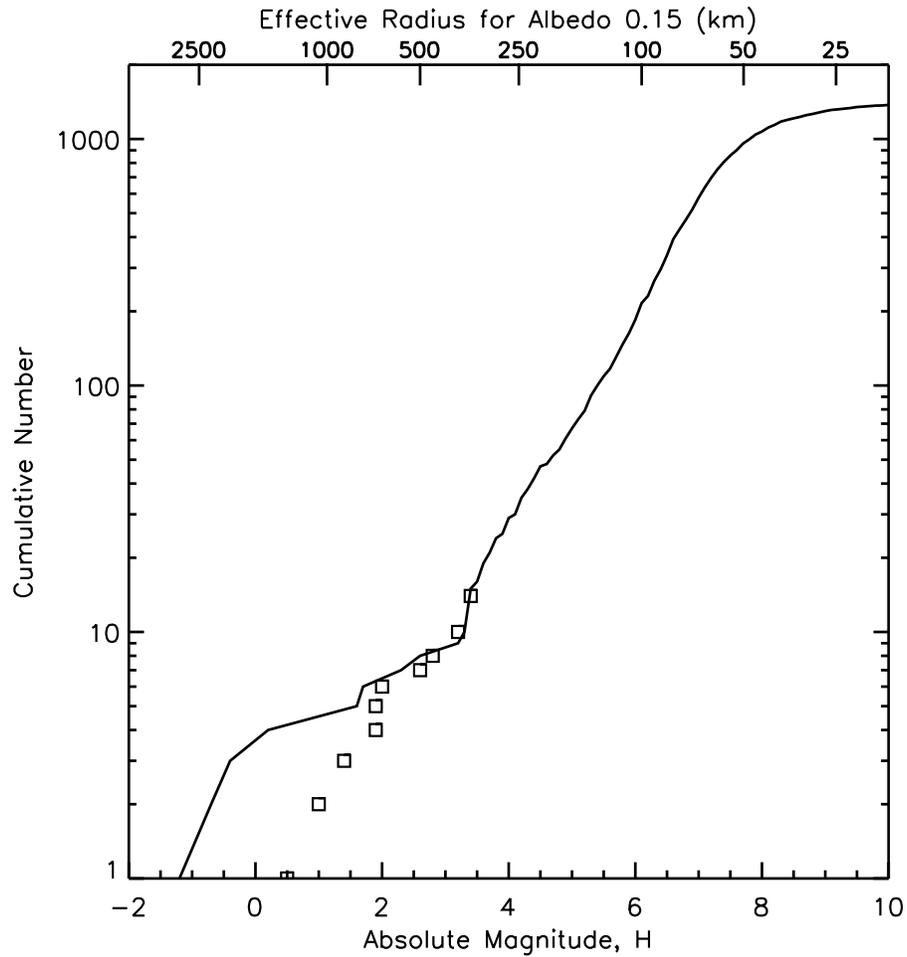}}
\caption{The absolute magnitude versus the cumulative number of all
  known trans-Neptunian objects (solid line).  The absolute magnitudes
  for the largest objects appear overly bright since these objects
  have much higher albedos than most smaller KBOs.  Squares show the
  absolute magnitudes that the largest KBOs would have if their
  albedos were 0.15 and not around 0.7 as has been found for Eris,
  Pluto, Makemake and Haumea.  Squares also show the absolute
  magnitudes the moderately sized KBOs would have if their albedos
  were not around 0.25 but 0.15 for Sedna, 2007 OR10, Orcus, and
  Quaoar (Stansberry et al. 2008).}
  \label{fig:KBOcumH} 
\end{figure}

Figure~\ref{fig:KBOcumH} shows the cumulative number of all known TNOs
versus their absolute magnitude, H.  Objects with absolute magnitudes
$H > 7$ mags appear to have a roll-over in their size distribution
because of detection biases.  The largest KBOs with $H<3$ mags do not
follow the simple power-law found for the objects with $3<H<7$ mags
(Brown 2008).  The largest KBOs have been found to have preferentially
higher albedos, likely because of atmosphere effects and surface
activity that keep the surfaces young and bright (Jewitt and Luu 2004;
Lykawka and Mukai 2005; Schaller and Brown 2007; Stansberry et
al. 2008; Desch et al. 2009).  The absolute magnitudes the largest
KBOs would have if they had a more typical albedo of 0.15 (Brown and
Trujillo 2004; Brown et al. 2006; Stansberry et al. 2008) are shown by
squares in Figure~\ref{fig:KBOcumH}.  The squares in
Figure~\ref{fig:KBOcumH} fit a simple power-law for all objects with
$H<7$ magnitudes ($r < 60$ km assuming 0.15 albedo).

\begin{figure}
\epsscale{0.4}
\centerline{\includegraphics[angle=0,totalheight=0.6\textheight]{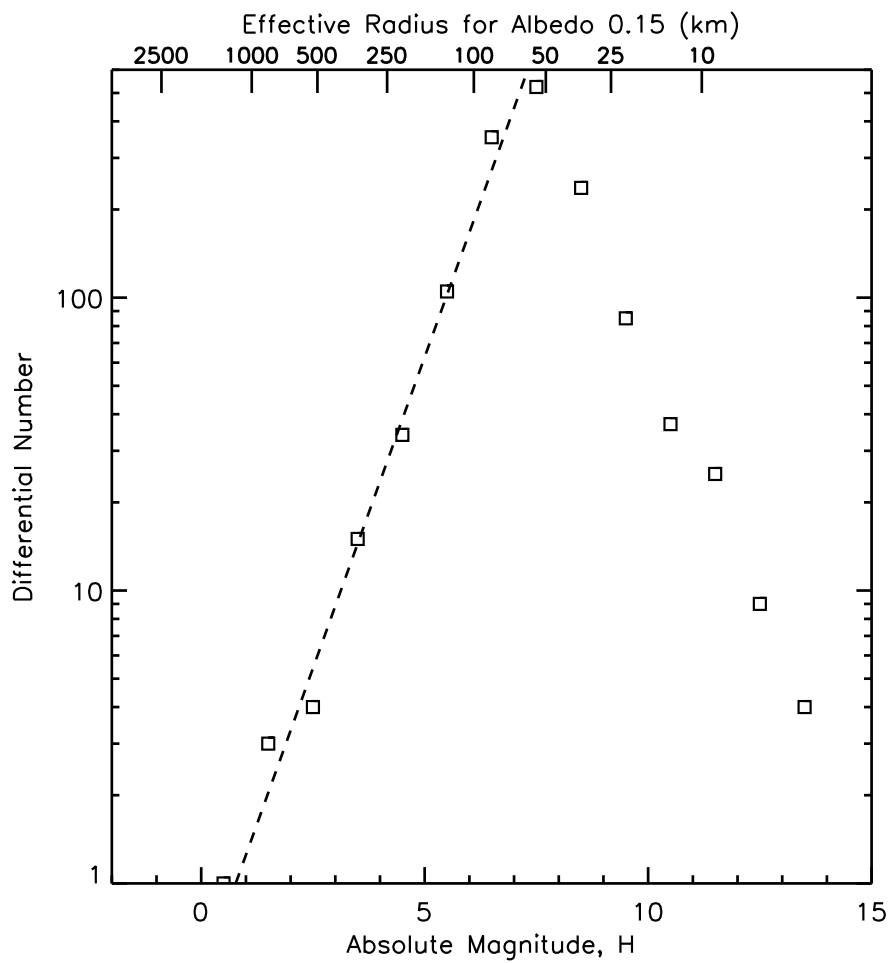}}
\caption{The absolute magnitude versus the differential number of
  known trans-Neptunian objects.  Objects are binned in 1 magnitude
  bins.  The largest few objects have had their absolute magnitudes
  adjusted fainter as in Figure~\ref{fig:KBOcumH} to account for their
  higher albedos compared to the smaller objects.  The dashed line
  shows the best fit to the largest objects.  It is apparent that the
  Kuiper Belt is nearly complete to about an absolute magnitude of
  around 5-6 mags after which a turnover shows significant
  incompleteness.}
\label{fig:KBOdiffH} 
\end{figure}

The points in a cumulative distribution are heavily correlated with
one another, tending to give excess weight to the faint end of the
distribution.  A differential distribution does not suffer from this
problem.  Figure~\ref{fig:KBOdiffH} shows the differential number of
all known TNOs versus their absolute magnitude, where, like in
Figure~\ref{fig:KBOcumH}, the largest few objects have had their
absolute magnitudes adjusted for their abnormally high albedos
compared to smaller objects.  It is clear there is a turnover around
an absolute magnitude of 7 mags ($r \sim 60$ km) showing observational
bias beyond this magnitude.  The best fit power-law for the
differential points finds $q=3.0\pm 0.5$ for $H<7$ magnitudes, where
$n(r)dr \propto r^{-q}dr$ is the differential power-law radius
distribution with $n(r)dr$ describing the number of TNOs with radii in
the range $r$ to $r+dr$.  This is slightly lower than most previous
fits ($q\sim 4$) that were more heavily dependent on fainter (smaller)
objects (Jewitt et al. 1998; Trujillo et al. 2001; Petit et al. 2008;
Fraser et al. 2008; Fuentes and Holman 2008; Fraser and Kavelaars
2009; Fuentes et al. 2009).  As the scattered disk and Sedna
populations are not close to completion on the large end ($H<4.5$
mags), including such objects (Eris, 2007 OR10 and Sedna), as done
here, likely results in a shallower measured slope.  The size
distributions of individual dynamical classes are likely more
informative (see section 4.2.2).  There are no obvious discontinuities
at the large end of the KBO size distribution when including all
dynamical classes of TNOs (Figures~\ref{fig:KBOcumH}
and~\ref{fig:KBOdiffH}).

\subsubsection{Dwarf Planets}

A dwarf planet is defined by the International Astronomical Union
(IAU) as an object that is in hydrostatic equilibrium and has not
cleared the neighborhood around its heliocentric orbit of other
similarly sized objects.  Though the dwarf planet definition is
imprecise, it is clear that Ceres in the main asteroid belt as well as
Pluto and Eris in the outer solar system are bonafide dwarf planets.
Makemake and Haumea are also likely dwarf planets as are the next
largest bodies in the outer solar system such as Sedna, 2007 OR10,
Orcus and Quaoar.  Though the lower size limit of an object in
hydrostatic equilibrium is not well defined, Lineweaver and Norman
(2010) suggest it could be as small as 200 km in radius for an icy
body in the outer solar system.  This would put tens more objects in
the outer solar system into the dwarf planet category, including three
objects discovered in this survey (Table 1: 2010 EK139, 2010 KZ39 and
2010 FX86).  The actual sizes and shapes of these bodies are not well
known to date and will depend heavily on their albedos and
compositions.  Further detailed observations are required to determine
the true sizes and shapes of the new discoveries.

\begin{figure}
\epsscale{0.4}
\centerline{\includegraphics[angle=90,totalheight=0.6\textheight]{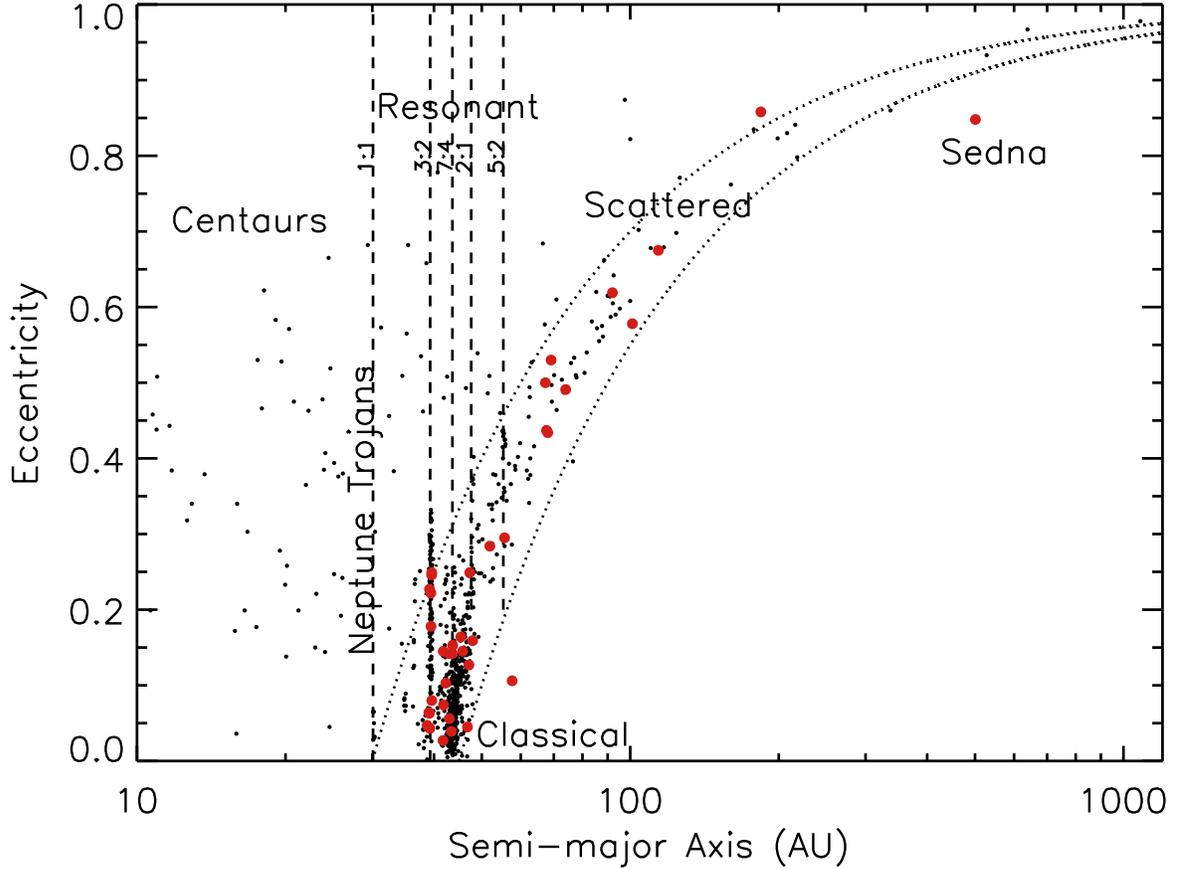}}
\caption{The semi-major axis versus eccentricity of multi-opposition
  trans-Neptunian objects.  Large circles represent TNOs with
  $H\leq 4.5$ mags.  This figure shows several distinct dynamical KBO
  populations. Vertical dashed lines show the main resonances with
  Neptune as well as the Neptune Trojans in the $1:1$ resonance.
  Scattered disk objects have perihelia $30 \lesssim peri \lesssim 45$
  AU as shown between the dashed lines.  Classical objects are
  in the lower center portion of the figure and include the the Main
  Kuiper Belt (MKB) with its high and low inclination populations.
  There also appears to be a high and low eccentricity population of
  large objects.  An edge near 50 AU can clearly be seen for low
  eccentricity objects.  Centaurs are on unstable orbits between the
  giant planets.  Sedna stands out as being significantly below the
  perihelion line shown at 40 AU demonstrating its decoupled influence
  from Neptune unlike the scattered disk objects.}
\label{fig:kboea2011} 
\end{figure}

With most of the biggest Kuiper Belt objects likely known, it is
interesting to compare where the largest ($H\leq 4.5$ mags) objects
reside dynamically in the Kuiper Belt (Figure~\ref{fig:kboea2011}).
At least one of the largest objects can be found in most of the TNO
dynamical populations (Tables 3 and 4).  The scattered disk population
(Gomes et al. 2008) has Eris and 2007 OR10, while Sedna is in its own
dynamical class (Morbidelli and Levison 2004; Gladman and Chan 2006)
that resides significantly beyond the Kuiper Belt edge (Trujillo and
Brown 2001; Allen et al. 2001).  The high inclination classical Kuiper
belt (Gomes 2003) has several large objects including Makemake,
Haumea, Varuna and (278361) 2007 JJ43.  Even the low inclination
classical Kuiper belt population, generally known for its smaller
sized objects (Levison and Stern 2001), appears to have Quaoar.
Further confirming Quaoar's status as a low inclination Kuiper belt
object is Quaoar's ultra-red surface (Jewitt and Luu 2004), which is a
characteristic generally associated with the low inclination classical
Kuiper Belt (Tegler and Romanishin 2000; Trujillo and Brown 2002;
Stern 2002; Doressoundiram et al. 2008; Peixinho et al. 2008).

The actual number of Pluto sized bodies is now known (Table 3).
Previous authors have argued that the Kuiper Belt likely lost a
substantial amount of its mass through collisional grinding and
dynamical interactions with the planets (Kenyon and Luu 1999; Levison
et al. 2008; Morbidelli et al. 2008; Stewart and Leinhardt 2009).
Observationally, many more objects appear to be required in order to
produce the observed angular momentum of the largest KBOs (Jewitt and
Sheppard 2002; Rabinowitz et al. 2006) and binaries (Noll et
al. 2008).  Detailed simulations show that Kuiper Belt formation is
possible with only the small number of Pluto sized objects observed
(Kenyon and Bromley 2008; Schlichting and Sari 2011).  A significant
number of Pluto sized objects likely exist in the populations beyond
100 AU such as the Sedna types and Oort cloud objects, which are
currently too faint to be efficiently detected to date.  It is
important to determine if the Pluto sized objects formed in the Kuiper
Belt as we see it today or if they originated much closer to the Sun
and were later transported to their current orbits.

\subsubsection{TNO Population Ratios}

\begin{figure}
\epsscale{0.4}
\centerline{\includegraphics[angle=0,totalheight=0.6\textheight]{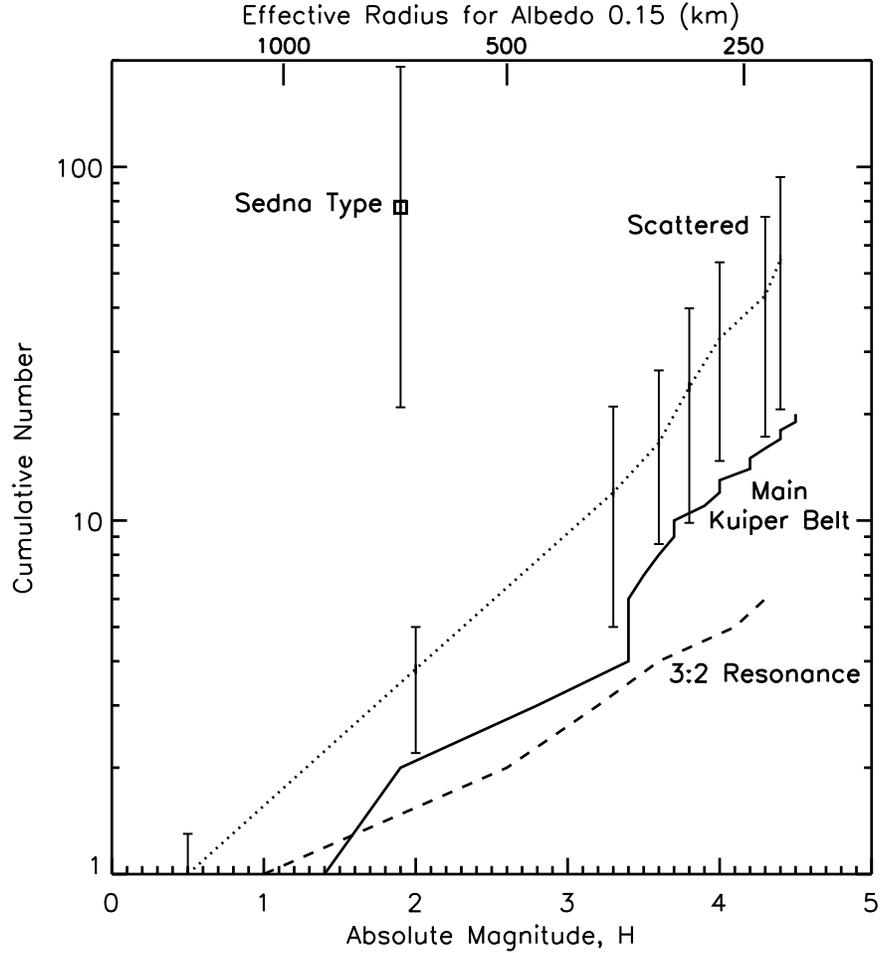}}
\caption{The absolute magnitude versus the cumulative number of
  objects for the various completeness limits (see Table 4) of the
  dynamical classes in the Kuiper Belt.  The largest few objects have
  had their absolute magnitudes adjusted as in
  Figure~\ref{fig:KBOcumH} to account for their higher albedos
  compared to the smaller objects.  The Sedna type (square) and
  scattered disk objects (dotted line) have significant error bars as
  these populations are not complete for even the largest objects.
  Poisson statistics were used to extrapolate the total scattered disk
  and Sedna populations from the known objects brighter than 21st
  magnitude using the amount of time the objects would be detectable
  in their eccentric orbits (Table 4).  The scattered population is
  likely larger than either the main Kuiper Belt (solid line) or the
  3:2 resonance population (dashed line).  The Sedna type population
  appears to be the largest of all the populations by a factor of ten
  or more.  The 3:2 resonance population has a shallower size
  distribution slope ($q=2.2\pm 0.5$) than the other populations
  ($q=3.3\pm 0.7$).}
\label{fig:KBO32} 
\end{figure}

On the large size end ($H \lesssim 4.5$ mags), the ratio of the
(Plutinos):(Main Kuiper Belt):(Scattered Disk):(Sedna Types) was found
to be $(1):(2.6):(7\pm3):(75\pm_{-55}^{+115})$, respectively (Table
4).  Thus the Sedna population could be the dominant observed small
body population for dwarf sized planets (Figure~\ref{fig:KBO32}).  The
scattered disk population is likely bigger than the main Kuiper Belt
(MKB) population by a factor of a few.  The Plutino population is
smaller by a factor of a few compared to the main Kuiper Belt.  Both
the scattered disk and main Kuiper Belt populations on the large end
of the size distribution ($H \lesssim 4.5$ mags) are consistent with
$q = 3.3\pm0.7$ while the Plutino population appears significantly
shallower than this with $q = 2.2\pm0.5$.  The scattered disk
population size determined from the largest objects ($H<4.5$ mags) is
consistent with Trujillo et al. (2001) estimated from smaller objects
in the scattered disk when using a $q\sim 3.3$ size distribution.

\subsubsection{The Main Kuiper Belt}

\begin{figure}
\epsscale{0.4}
\centerline{\includegraphics[angle=0,totalheight=0.6\textheight]{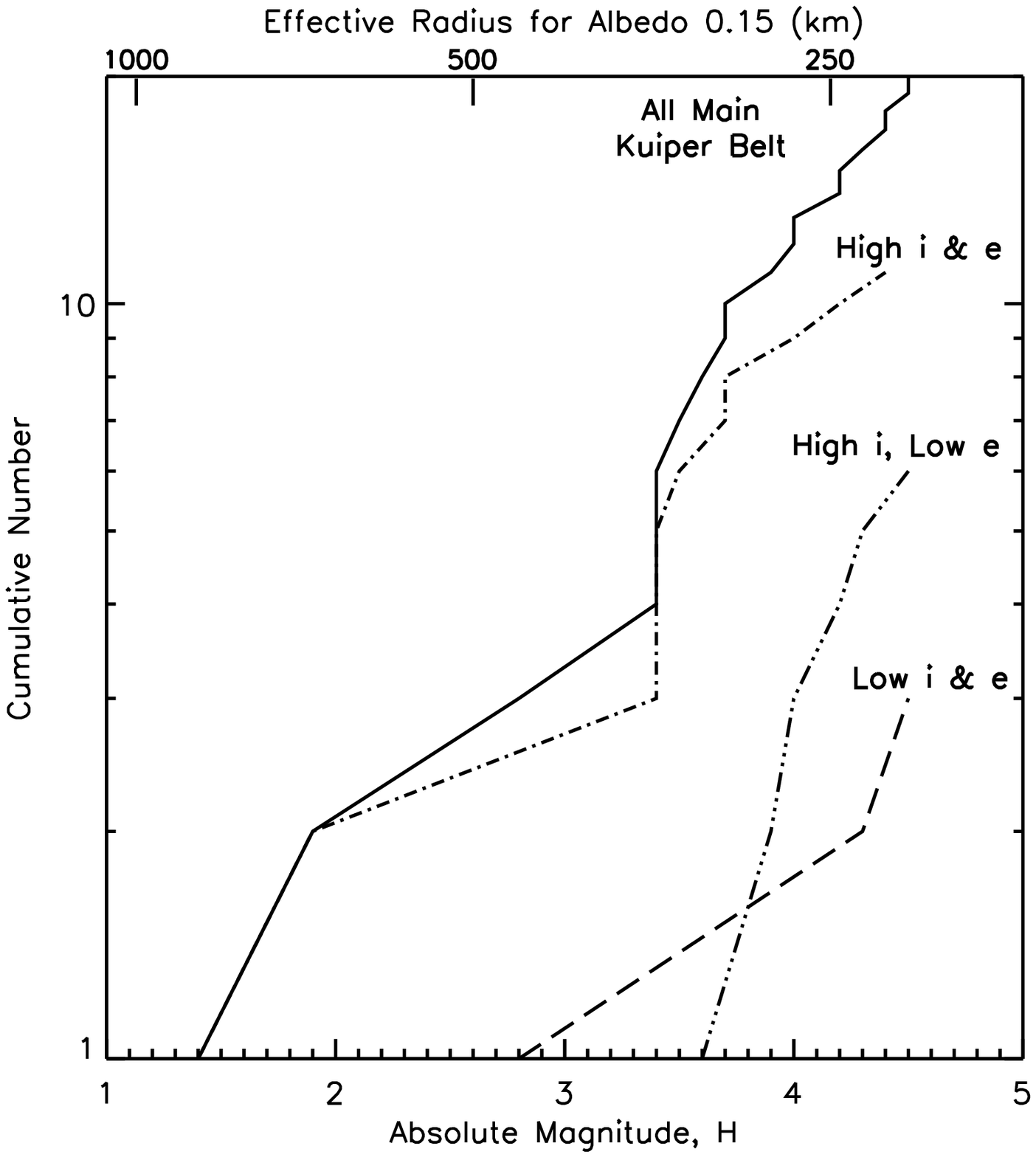}}
\caption{The absolute magnitude versus the cumulative number of
  objects in the main Kuiper Belt (solid line) and its subcategories.
  The largest few objects have had their absolute magnitudes adjusted
  as in Figure~\ref{fig:KBOcumH} to account for their higher albedos
  compared to the smaller objects.  The high inclination ($i>10$
  degrees) and eccentricity ($e>0.13$) objects (dotted dashed line)
  dominate the main Kuiper Belt on the large end.  There appears to be
  a sizable group of main Kuiper Belt objects that have high
  inclinations but low eccentricities ($e<0.07$) (triple dotted dashed
  line).  These high $i$ and low $e$ objects could be related to
  either of the other subcategories in this figure.  Large low
  inclination and low eccentricity objects (long dashed line) are very
  rare with only Quaoar being brighter than an absolute magnitude of 4.2.}
\label{fig:KBOmkb} 
\end{figure}

The main Kuiper Belt ($39 < a < 48$ AU) appears to be divided into
three distinct dynamical classes (Figure~\ref{fig:KBOmkb}).  The high
inclination and low inclination (``cold'') classical classes have been
suggested for a decade, with the largest objects preferentially in
high inclination orbits (Levison and Stern 2001; Brown 2001).  When
plotting only the largest few objects ($H<4.5$ mags), there appears to
also be both low eccentricity and higher eccentricity classes
(Figure~\ref{fig:kboea2011}).  All three of the low inclination
objects ($i<10$ degs) with $H<4.5$ mags have low eccentricities
($e<0.05$).  The high inclination objects with $H<4.5$ mags in the
main Kuiper Belt appear to have either low eccentricities ($0.03 < e <
0.07$; 6 observed) or significantly higher eccentricities ($0.13 < e <
0.16$; 8 observed).  Only one of the twenty main Kuiper Belt objects
with $H<4.5$ mags has an eccentricity between these two ranges
(Salacia (120347) 2004 SB60 which has $e=0.10$) while two others have
slightly higher eccentricities (Haumea with $e=0.20$ and (230965) 2004
XA192 with $e=0.25$).

The Hartigan and Hartigan (1985) dip test for bimodality shows a
strong bimodality in eccentricity when including all main Kuiper Belt
objects with $H<4.5$ mags except for the interesting binary object
Salacia (these nineteen objects give a dip statistic of 0.145 which
corresponds to a confidence of 0.997 for bimodality, a 3 sigma
result).  Including Salacia in the dip test gives a less significant
result of only 0.990 confidence in bimodality, or slightly less than 3
sigma.  Including smaller main Kuiper Belt objects decreases the
bimodality significance even further.  If real, the low versus higher
eccentricity populations of highly inclined large objects could have
different origins, such as forming in different regions of the solar
system or originally from different scattering events during the
migration of the planets.

The largest bodies ($H<4.5$ mags) are too few for meaningful
statistics, but it appears that the high inclination main Kuiper Belt
does not have a significantly shallower power-law distribution than
the low inclination population, as was found for the smaller objects
of these two populations by Fraser et al. (2010)
(Figure~\ref{fig:KBOmkb}).  There is a possible deficiency of objects
in the main Kuiper Belt between $2.5 < H < 3.5$ magnitudes, but this
is likely not statistically significant as it is just small number
statistics.

\subsubsection{Resonance Populations}

A surprising result is the absence of large objects in all the main
Neptune resonance populations except the 3:2 resonance (see Gladman et
al. 2008 and Elliot et al. 2005 for resonance calculations as well as
the updated version of Elliot et al. 2005 kept by Marc Buie at
www.boulder.swri.edu/buie/kbo/astrom).  The Plutinos or 3:2 resonance
objects include some of the largest known KBOs such as Pluto, Orcus
and Ixion (Table 3) while the other observed heavily populated
resonances such as the 5:3, 7:4, 2:1, and 5:2 have no known large KBOs
(Table 4).  Any object in the Neptune resonances brighter than 21st
magnitude ($r \gtrsim 200$ km), would likely have been detected by now
(Figure~\ref{fig:distanceH} and Table 4).  The 5:2 has a few sizable
objects with absolute magnitudes of around 3.8 and 5.1 mags, (84522)
2002 TC302 and (26375) 1999 DE9, respectfully.  The largest 2:1 object
appears to be (119979) 2002 WC19 with an absolute magnitude of 5.1
mags.  None of the other resonances have any objects with absolute
magnitudes brighter than 5 mags.  2010 EK139, discovered in this
survey, appears to be one of the only known objects in the very
distant 7:2 resonance (based on orbit calculations from Marc Buie's
website at www.boulder.swri.edu/buie/kbo/astrom that has up to date
information first published in Elliot et al. 2005).

The relative populations of the various Neptune resonances are
currently not well constrained since observational biases make
discoveries easier in the closer 3:2 resonance (Jewitt et al. 1998;
Trujillo et al. 2001).  There is also a strong longitude and latitude
dependence on discovery of resonance populations (Chiang and Jordan
2002; Chiang et al. 2003).  Previous observational works have
suggested that the 2:1 resonance appears to have less objects than the
3:2 resonance (Jewitt et al. 1998; Chiang and Jordan 2002).  Numerical
simulations of resonance sweeping (Hahn and Malhotra 2005) have shown
that the main Neptune resonance populations relative to the 3:2
resonance population may be 2:1 (x2), 7:4 (x0.8), 5:3 (x0.6), 5:2
(x0.5).  These simulations suggest that the outer resonances should
have a factor of four more objects than the 3:2 resonance.  Thus, if 3
very large objects such as Pluto, Orcus and Ixion were found in the
3:2 resonance population, based on Poisson statistics, one would
expect $12\pm3.5$ objects of similar size in the other resonances.
This is not the case, and such a scenario can be rejected with 3.5
sigma confidence, so either the outer resonances are significantly
less populated than the 3:2 resonance or the outer resonance bodies
have a different size distribution than the 3:2 resonance.  It is
likely that the 3:2 resonance is populated by objects that formed
significantly closer to the Sun than the outer resonances.  Objects
forming closer to the Sun would likely accrete more material in a
shorter amount of time allowing them to become larger before they were
captured in the Neptune resonances.  With the Kuiper Belt nearly
complete to 21st magnitude, it is unlikely that a planet larger than
Mercury within a few 100 AU currently exists.  Lykawka and Mukai
(2008) suggested such a planet could have disrupted the outer
resonance populations.  It is still possible that a close stellar
encounter or now defunct outer planet could have disrupted or depleted
the outer resonances early in the solar system's history.

\section{Summary}

The OGLE Carnegie Kuiper belt Survey (OCKS) is one of the first
southern sky and galactic plane surveys for bright outer solar system
objects.  Eighteen bright Trans-Neptunian objects were discovered,
including some of the most southern outer solar system objects ever
detected as well as the intrinsically brightest solar system objects
discovered in several years (2010 EK139 with $H=3.8$ and 2010 KZ39
with $H=3.9$ mags).

1) A total of 2500 square degrees was searched in the survey.  About
2200 square degrees of the survey was south of declination -25
degrees, where northern KBO surveys cannot efficiently observe.  The
surveyed area includes almost all of the southern sky within about 20
degrees of the ecliptic.  Another 300 square degrees of sky was
surveyed in the northern galactic plane near the ecliptic using
optimal image subtraction techniques to remove the stellar background.

2) The survey obtained a limiting R-band magnitude of 21.6 during
optimal observing conditions using the 1.3 meter Warsaw telescope at
Las Campanas observatory in Chile.  In moderate seeing the survey
limit was 21.2 magnitudes in the R-band.  During bad seeing conditions
the survey was not performed.

3) Kuiper Belt surveys are now nearly complete to about 21st magnitude
in the R-band.  The corresponding size of an object at 21st magnitude
depends on the distance and albedo of the object.  At 30 AU 21st
magnitude corresponds to about $H=6.6$ mags while at 50 AU $H= 4.4$
mags, which when assuming a moderate albedo of $\rho_{R}=0.15$
correspond to radii of 80 km and 225 km respectively.  Through looking
at the cumulative luminosity function of the Kuiper Belt objects,
significant incompleteness in the main Kuiper Belt probably starts
around a radius of 100 km ($H \sim 6$ mags) and becomes drastic around
a radius of 60 km ($H \sim 7$ mags.).

4) For the largest objects ($H \lesssim 4.5$ mag), the ratio of the
population sizes for the various dynamical reservoirs in the outer
solar system were found to be
$(1):(2.6):(7\pm3):(75\pm_{-55}^{+115})$, for the (Plutinos):(Main
Kuiper Belt):(Scattered Disk):(Sedna Types), respectively.  Thus the
scattered disk population is likely a few times larger than the main
Kuiper Belt population and several times larger than the Plutino
population.  The Sedna type population likely is the biggest of all
the observed outer solar system reservoirs but remains largely unknown
because of the strong observational bias against finding very distant
objects.

5) Beyond the Kuiper Belt edge, at a few hundred AU or so, there could
easily be more Pluto, Mercury or even larger sized objects in
Sedna-like orbits.  No new Sedna-like objects were detected even
though the survey was sensitive to objects up to about 300 AU.  Sedna
is likely one of the larger and thus one of the brighter members of
its population.  Any further Sedna-like object detections will likely
require significantly fainter magnitudes while still covering large
areas of sky.  Pan-Starrs has a chance to detect some Sedna like
objects since it will survey large areas of sky to around a magnitude
fainter than this survey, but LSST will be needed to find significant
numbers of Sedna like objects since sensitivity and large areas of sky
are needed to probe this distant, faint population.

6) All the major populated dynamical reservoirs in the Kuiper Belt,
including the scattered disk, high inclination classical belt, low
inclination classical belt (Quaoar), Sedna and the Plutinos are
occupied by dwarf planet sized objects.  Only the well-populated outer
Neptune mean motion resonances such as the 2:1, 7:4, 5:2, and 5:3 are
not occupied by a dwarf planet sized object.  Any dwarf planet in these
outer resonances would likely have been found to date, suggesting the
outer resonances are occupied by a different mix of objects than the
3:2 resonance population or are significantly depleted in objects
relative to the 3:2 resonance.

7) The scattered disk and main Kuiper Belt were found to have a
power-law size distribution of $q=3.3\pm0.7$ for the largest few
objects ($H<4.5$ mags), while the Plutino population has a shallower
slope of $q=2.2\pm0.5$.  The high and low inclination main Kuiper Belt
populations appear to have similar slopes in their size distributions.

8) The main Kuiper Belt could have three distinct dynamical classes:
(1) low inclination with low eccentricity ($e<0.05$), (2) high
inclination with low eccentricity ($e<0.07$), and (3) high inclination
with higher eccentricities ($e>0.13$).

\section*{Acknowledgments}

The OGLE project has received funding from the European Research
Council under the European Community's Seventh Framework Programme
(FP7/2007-2013) / ERC grant agreement no. 246678 to AU.  C.T. was
supported by the Gemini Observatory, which is operated by the
Association of Universities for Research in Astronomy, Inc., on behalf
of the international Gemini partnership of Argentina, Australia,
Brazil, Canada, Chile, the United Kingdom, and the United States of
America.

\newpage



\begin{center}
\begin{deluxetable}{lccccccc}
\tablenum{1}
\tablewidth{5 in}
\tablecaption{New outer Solar System objects discovered in this survey}
\tablecolumns{8}
\tablehead{
\colhead{Name} & \colhead{H} & \colhead{$m_{R}$} & \colhead{$a$}  & \colhead{$e$} & \colhead{$i$} & \colhead{$R$} & \colhead{$r$} \\ \colhead{} & \colhead{(mag)} & \colhead{(mag)} &\colhead{(AU)} & \colhead{} & \colhead{(deg)} & \colhead{(AU)} & \colhead{(km)} }  
\startdata
2010 EK$_{139}$   &    3.8  &  19.5  &  69.1  &  0.53  &  29.5  &  40.5  &  310\tablenotemark{a}    \nl
2010 KZ$_{39}$    &    3.9  &  20.1  &  45.8  &  0.15  &  26.1  &  46.3  &  300\tablenotemark{a}    \nl
2010 FX$_{86}$    &    4.3  &  20.7  &  47.0  &  0.08  &  25.2  &  46.8  &  230\tablenotemark{a}    \nl
2010 EL$_{139}$   &    5.0  &  20.1  &  39.2  &  0.07  &  23.0  &  36.6  &  190    \nl
2010 HE$_{79}$    &    5.1  &  19.8  &  39.3  &  0.20  &  15.7  &  34.9  &  180    \nl
2010 PU$_{75}$    &    5.3  &  20.9  &  43.4  &  0.08  &  10.2  &  40.0  &  150    \nl
2010 JK$_{124}$   &    5.4  &  21.2  &  39.7  &  0.09  &  15.6  &  40.3  &  140    \nl
2009 MF$_{10}$    &    6.0  &  21.1  &  57.5  &  0.52  &  26.1  &  36.1  &  120     \nl
2010 HD$_{112}$   &    6.5  &  22.2  &  44.5  &  0.03  &   3.9  &  43.1  &  100     \nl
2010 JJ$_{124}$   &    6.6  &  20.1  &  83.0  &  0.72  &  37.8  &  24.1  &  80     \nl
2009 MG$_{10}$    &    7.0  &  21.7  &  47.5  &  0.34  &  19.9  &  32.8  &  70     \nl
2010 HG$_{109}$   &    7.3  &  21.7  &  39.8  &  0.23  &  29.2  &  30.5  &  60     \nl
2010 HU$_{113}$   &    7.4  &  22.1  &  36.2  &  0.03  &  11.3  &  35.3  &  60     \nl
2009 ME$_{10}$    &    7.5  &  21.0  &  27.8  &  0.18  &  14.7  &  23.1  &  50     \nl
\enddata
\tablecomments{
Orbital elements are from the Minor Planet Center and are the
semimajor axis ($a$), inclination ($i$), and eccentricity ($e$).  The
radii ($r$) of the new objects were determined assuming an albedo of
0.15 and using the equation, $r = (2.25\times 10^{16} R^{2} \Delta
^{2} / p_{R}\phi (0))^{1/2} 10^{0.2(m_{\odot} - m_{R})}$ where $R$ is
the heliocentric distance in AU, $\Delta$ is the geocentric distance
in AU, $m_{\odot}$ is the apparent red magnitude of the sun ($-27.1$),
$p_{R}$ is the red geometric albedo, $m_{R}$ is the apparent red
magnitude of the object and $\phi (0) = 1$ is the phase function at
opposition.  H is the absolute magnitude of the object.}
\tablenotetext{a}{These objects could be labeled as dwarf planets since their radii are larger than 200 km assuming a moderate or lower albedo.}
\end{deluxetable}
\end{center}


\newpage



\begin{center}
\begin{deluxetable}{lccccccc}
\tablenum{2}
\tablewidth{5 in}
\tablecaption{Known KBOs and Centaurs detected in this survey}
\tablecolumns{8}
\tablehead{
\colhead{Name} & \colhead{H} & \colhead{$m_{R}$} & \colhead{$a$}  & \colhead{$e$} & \colhead{$i$} & \colhead{$R$} & \colhead{$r$} \\ \colhead{} & \colhead{(mag)} & \colhead{(mag)} &\colhead{(AU)} & \colhead{} & \colhead{(deg)} & \colhead{(AU)} & \colhead{(km)} }  
\startdata
(134340) Pluto          &   -0.7  &  13.6  &  39.6  &  0.25  &  17.1  &  31.8  &  1150   \nl
2007 JJ$_{43}$           &    3.2  &  19.4  &  48.0  &  0.16  &  12.1  &  41.7  &  350    \nl 
(10199) Chariklo        &    6.4  &  17.5  &  15.8  &  0.17  &  23.4  &  13.8  &  100    \nl
(55576) Amycus          &    7.8  &  19.6  &  25.0  &  0.39  &  13.3  &  16.8  &  50     \nl
\enddata
\tablecomments{See Table 1 for comments and definitions.}
\end{deluxetable}
\end{center}


\newpage



\begin{center}
\begin{deluxetable}{lccccc}
\tablenum{3}
\tablewidth{5.5 in}
\tablecaption{Ten Intrinsically Brightest TNOs}
\tablecolumns{6}
\tablehead{
\colhead{Name} & \colhead{H} & \colhead{$a$}  & \colhead{$e$} & \colhead{$i$} & \colhead{Class}  \\ \colhead{} & \colhead{(mag)}  &\colhead{(AU)} & \colhead{} & \colhead{(deg)} & \colhead{} }  
\startdata
(136199) Eris         &   -1.2  & 68.0  & 0.43 &  43.9 & Scattered  \\
(134340) Pluto        &   -0.7  & 39.7  & 0.25 &  17.1 & 3:2 Resonance  \\
(136472) Makemake     &   -0.3  & 45.4  & 0.16 &  29.0 & High $i$ Classical  \\
(136108) Haumea       &    0.2  & 43.0  & 0.20 &  28.2 & High $i$ Classical  \\
 (90377) Sedna        &    1.6  &   510   & 0.85 &  11.9 & Sedna  \\
(225088) 2007 OR10    &    1.9  & 67.3  & 0.50 &  30.7 & Scattered  \\
 (90482) Orcus        &    2.3  & 39.2  & 0.23 &  20.6 & 3:2 Resonance  \\
 (50000) Quaoar       &    2.5  & 43.5  & 0.04 &   8.0 & Low $i$ Classical  \\
 (28978) Ixion        &    3.2  & 39.6  & 0.25 &  19.6 & 3:2 Resonance  \\  
\enddata
\tablecomments{The orbital elements are from the Minor Planet Center and are the semimajor axis ($a$), inclination ($i$), and eccentricity ($e$).  H is the absolute magnitude and Class is the dynamical classification of the object.}
\end{deluxetable}
\end{center}


\newpage



\begin{center}
\begin{deluxetable}{lcccc}
\tablenum{4}
\tablewidth{5.5 in}
\tablecaption{Bright Kuiper Belt Population Statistics}
\tablecolumns{5}
\tablehead{
\colhead{Class} & \colhead{$H_{comp}$} & \colhead{$r_{comp}$}  & \colhead{$N$} & \colhead{Pop Ratio}  \\ \colhead{} & \colhead{(mag)}  &\colhead{(km)} & \colhead{} & \colhead{($N/N_{3:2}$)}}  
\startdata
3:2                      &   4.5   &   210   &  6              &   1         \\
5:3                      &   4.2   &   250   &  0              &   $0/5$    \\
7:4                      &   4.1   &   260   &  0              &   $0/5$    \\
2:1                      &   3.3   &   380   &  0              &   $0/3$    \\
5:2                      &   2.5   &   550   &  0              &   $0/2$    \\
Main Kuiper Belt (MKB)   &   4.1   &   260   &  13\tablenotemark{a} &   $13/5$\tablenotemark{a}        \\
Scattered Disk           &   4.1\tablenotemark{b}    &   N/A   &  $35\pm15$\tablenotemark{b}   &  $35\pm15/5$        \\
Sedna Type               &   1.6\tablenotemark{c}    &   N/A   &  $75_{-55}^{+115}$\tablenotemark{c} & $75_{-55}^{+115}/1$  \\
MKB Low i \& e           &   4.6   &   200   &  3              &   $3/6$   \\
MKB High i, All e        &   4.1   &   260   &  11             &   $11/5$     \\
MKB High i \& e          &   4.1   &   260   &  8              &   $8/5$     \\
MKB High i \& Low e      &   4.6   &   200   &  6              &   $6/6$     \\
\enddata
\tablecomments{$H_{comp}$ is the absolute magnitude completion limit for the particular dynamical class while $r_{comp}$ is the radius completion limit assuming an albedo of 0.15.  $N$ is the number of objects known within each class with an aboslute magnitude equal to or brighter than $H_{comp}$.  The Pop Ratio is the population number ratio of each dynamical class relative to the 3:2 resonance number population at the $H_{comp}$ of that particular dynamical class (i.e. $N/N_{3:2}$).}
\tablenotetext{a}{None of the Haumea family members, except for Haumea itself, are included.  The Haumea family members are likely pieces of Haumea and have very high albedos unlike most of the other moderately sized objects with absolute magnitudes around 3 or 4 (see Raggozine et al. 2007; Trujillo et al. 2011).}
\tablenotetext{b}{Since the scattered disk objects spend most of their time near aphelion, which can be up to a few hundred AU, the absolute magnitude completion number here is for objects currently within about 50 AU of the Sun.  The total number of possible scattered disk objects with absolute magnitdue brighter than this was determined by taking the number of known objects of this brightness or brighter and a Poisson probability statistic of how many more are currently unobservable in the distant solar system based on the percent of time the known objects would be brighter than 21st magnitude in their orbit.}
\tablenotetext{c}{Like the scattered disk objects, Sedna is only brighter than 21st magnitude near perihelion.  Thus for most of Sedna's orbit it would not be detected by the current large area surveys.  To account for this, a Poisson probability statistic of how many more Sedna type objects of similar size are unobservable in the distant solar system was determined based on Sedna's orbit.}
\end{deluxetable}
\end{center}


\newpage

\newpage

\newpage

%

\newpage

\newpage

\newpage

\newpage

\newpage

\newpage

\newpage

\newpage

\newpage

\begin{references}

\reference{Ala98} Alard, C. 2000, A\&AS, 144, 363.

\reference{Ala98} Alard, C. and Lupton, R. 1998, ApJ, 503, 325.

\reference{All01} Allen, L., Bernstein, G. and Malhotra, R. 2001, ApJ, 549, L241.

\reference{Bar08} Barucci, M., Brown, M., Emery, J. and Merlin, F. 2008, in The Solar System Beyond Neptune, ed. M. Barucci, H. Boehnhardt, D. Cruikshank and A. Morbidelli (Tucson: Univ of Arizona Press), 143-160. 

\reference{Bro01} Brown, M. 2001, AJ, 121, 2804.

\reference{Bro04} Brown, M. and Trujillo, C. 2004, AJ, 127, 2413.

\reference{Bro04} Brown, M., Trujillo, C. and Rabinowitz, D. 2004, ApJ, 617, 645.

\reference{Bro05} Brown, M., Trujillo, C. and Rabinowitz, D. 2005, ApJ, 635, L97.

\reference{Bro06} Brown, M. Schaller, E., Roe, H., Rabinowitz, D. and
Trujillo, C. 2006, ApJ, 643, L61.

\reference{Bro08} Brown, M. 2008, in The Solar System Beyond Neptune, ed. M. Barucci, H. Boehnhardt, D. Cruikshank and A. Morbidelli (Tucson: Univ of Arizona Press), 335-344.

\reference{Chi02} Chiang, E. and Jordan, A. 2002, AJ, 124, 3430.

\reference{Chi03} Chiang, E., Jordan, A., Millis, R. et al. 2003, AJ, 126, 430.

\reference{Cuz10} Cuzzi, J., Hogan, R., and Bottke, W. 2010, Icarus, 208, 518.

\reference{Des09} Desch, S., Cook, J., Doggett, T. and Porter, S. 2009, Icarus, 202, 694.

\reference{Dor08} Doressoundiram, A., Boehnhardt, H., Tegler, S. and Trujillo, C. 2008, in The Solar System Beyond Neptune, ed. M. Barucci, H. Boehnhardt, D. Cruikshank and A. Morbidelli (Tucson: Univ of Arizona Press), 91-104.
\reference{Dum07} Dumas et al. 2007, AA, 471, 331.

\reference{Ell03} Elliot, J. and Kern, S. 2003, EM\&P, 92, 375.

\reference{Ell05} Elliot, J., Kern, S., Clancy, K. et al. 2005, AJ, 129, 1117

\reference{Ell10} Elliot et al. 2010, Nature, 465, 897.

\reference{Fra08} Fraser, W. et al. 2008, Icarus, 195, 827.

\reference{Fra09} Fraser, W. \& Kavelaars, J. 2009, AJ, 137, 72.

\reference{Fra10} Fraser, W., Brown, M. \& Schwamb, M. 2010, 210, 944.

\reference{Fue08} Fuentes, C. \& Holman, M. 2008, AJ, 136, 83

\reference{Fue09} Fuentes, C., George, M., \& Holman, M. 2009, ApJ, 696, 91

\reference{Gla06} Gladman, B. and Chan, C. 2006, ApJ Lett., 643, L135.

\reference{Gla08} Gladman, B. Marsden, B. and VanLaerhoven, C. 2008, in The Solar System Beyond Neptune, ed. M. Barucci, H. Boehnhardt, D. Cruikshank and A. Morbidelli (Tucson: Univ of Arizona Press), 43-57.

\reference{Gom03} Gomes, R. 2003, Earth Moon Planets, 92, 29.

\reference{Gom05} Gomes, R., Levison, H., Tsiganis, K. and Morbidelli, A. 2005, Nature, 435, 466.

\reference{Gom08} Gomes, R., Fernandez, J., Gallardo, T. and Brunini, A. 2008, in The Solar System Beyond Neptune, ed. M. Barucci, H. Boehnhardt, D. Cruikshank and A. Morbidelli (Tucson: Univ of Arizona Press), 259-273.


\reference{Hah05} Hahn, J. and Malhotra, R. 2005, AJ, 130, 2392.

\reference{Har85} Hartigan, J. and Hartigan, P. 1985, The Annals of Statstics, 13, 70.

\reference{Ior09} Iorio, L. 2009, MNRAS, 400, 346.

\reference{Jew08} Jewitt, D., Luu, J., Trujillo, C. 1998, AJ, 115, 2125.

\reference{Jew02} Jewitt, D. and Sheppard, S. 2002, AJ, 123, 2110.

\reference{Jew04} Jewitt, D. and Luu, J. 2004, Nature, 432, 731.




\reference{Ken99b}  Kenyon, S. and Luu, J. 1999, AJ, 118, 1101.

\reference{Ken08}  Kenyon, S. and Bromley, B. 2008, ApJ Supplement, 179, 451.

\reference{Ken08} Kenyon, S., Bromely, B., O'Brien, D. and Davis, D. 2008, in The Solar System Beyond Neptune, ed. M. Barucci, H. Boehnhardt, D. Cruikshank and A. Morbidelli (Tucson: Univ of Arizona Press), 293-313.

\reference{Ken10} Kenyon, S and Bromley, B. 2010, ApJ Supplement, 188, 242.

\reference{Lev01} Levison, H. and Stern, S. A. 2001, AJ, 121, 1730.

\reference{Lev08} Levison, H., Morbidelli, A., Vanlaerhoven, C., Gomes, R., and Tsiganis, K. 2008, Icarus, 196, 258.

\reference{Lic06} Licandro, J. et al. 2006, AA, 445, 35.

\reference{Lin10} Lineweaver, C. and Norman, M. 2010, arXiv:1004.1091

\reference{Luu88} Luu, J. and Jewitt, D. 1988, AJ, 95, 1256.

\reference{Lyk05} Lykawka, P. S., and Mukai, T. Planetary and Space Science, 53, 1319.

\reference{Mal95} Malhotra, R. 1995, AJ, 110, 420.

\reference{Mel10} Melott, A. and Bambach, R. 2010, MNRAS, 407, L99.

\reference{Mor04} Morbidelli, A.and Levison, H. 2004, AJ, 128, 2564.

\reference{Mor08} Morbidelli, A., Levison, H. and Gomes, R.  2008, in The Solar System Beyond Neptune, ed. M. Barucci, H. Boehnhardt, D. Cruikshank and A. Morbidelli (Tucson: Univ of Arizona Press), 275-292.

\reference{Noll08} Noll, K., Grundy, W., Chiang, E., Margot, J. and Kern, S. 2008, in The Solar System Beyond Neptune, ed. M. Barucci, H. Boehnhardt, D. Cruikshank and A. Morbidelli (Tucson: Univ of Arizona Press), 345-363.

\reference{Pei08} Peixinho, N., Lacerda, P., and Jewitt, D. 2008, AJ, 136, 1837-1845.

\reference{Pet08} Petit, J., Kavelaars, J., Gladman, B. and Loredo, T. 2008, in The Solar System Beyond Neptune, ed. M. Barucci, H. Boehnhardt, D. Cruikshank and A. Morbidelli (Tucson: Univ of Arizona Press), 71-87. 

\reference{Rab06} Rabinowitz, D., Barkume, K., Brown, M., Roe, H., Schwartz, M., Tourtellotte, S. and Trujillo, C. 2006, ApJ, 639, 1238.

\reference{Rab07} Rabinowitz, D., Schaefer, B. and Tourtellotte, S. 2007, AJ, 133, 26.

\reference{Rab10} Rabinowitz, D. 2010, Poster presentation at the TNO 2010: Dynamical and Physical Properties of Trans-Neptunian Objects workshop in Philadelphi, PA, USA

\reference{Rag07} Ragozzine, D. and Brown, M. 2007, AJ, 134, 2160.

\reference{Sch07} Schaller, E. and Brown, M. 2007, ApJ, 659, L61.

\reference{Sch2011} Schlichting, H. and Sari, R. 2011, ApJ, 728, 68.

\reference{Sch09} Schwamb, M., Brown, M., and Rabinowitz, D. 2009, ApJ, 694, L45.

\reference{Sch10} Schwamb, M., Brown, M., Rabinowitz, D. and Ragozzine, D. 2010, ApJ, 720, 1691.

\reference{She07} Sheppard, S. 2007, AJ, 134, 787.

\reference{She10} Sheppard, S. and Trujillo, C. 2010, ApJ, 723, 233.

\reference{Sta08} Stansberry, J., Grundy, W., Brown, M., Cruikshank, D., Spencer, J., Trilling, D. and Margot, J. 2008, in The Solar System Beyond Neptune, ed. M. Barucci, H. Boehnhardt, D. Cruikshank and A. Morbidelli (Tucson: Univ of Arizona Press), 161-179.

\reference{Ste02} Stern, S. A. 2002, AJ, 124, 2297.

\reference{Ste09} Stewart, S. and Leinhardt, Z. 2009, ApJ Letters, 691, L133.

\reference{Teg00} Tegler, S. and Romanishin, W. 2000, Nature, 407, 979-981.

\reference{Tho11} Thomas-Osip, J., McCarthy, P., Prieto, G., Phillips, M., and Johns, M. 2011, arxiv:1101.2340


\reference{Tru98} Trujillo, C. and Jewitt, D. 1998, AJ, 115, 1680.

\reference{Tru00} Trujillo, C., Jewitt, D. and Luu, J. 2000, ApJ, 529, L103.

\reference{Tru01} Trujillo, C., Jewitt, D. and Luu, J. 2001, AJ, 122, 457.

\reference{Tru01b} Trujillo, C. and Brown, M. 2001, ApJ, 554, L95.

\reference{Tru02} Trujillo, C. and Brown, M. 2002, ApJ, 566, L125.

\reference{Tru03} Trujillo, C. and Brown, M. 2003, EM\&P, 92, 99.

\reference{Tru10} Trujillo, C., Sheppard, S. and Schaller, E. 2011, ApJ, 730, 105

\reference{Uda94} Udalski, A., Szymanski, M., Stanek, K. et al. 1994, Acta Astronomica, 44, 165.

\reference{Uda97} Udalski, A., Kubiak, M., and Szymanski, M. 1997, Acta Astronomica, 47, 319.

\reference{Uda03} Udalski, A. 2003, Acta Astronomica, 53, 291.

\reference{Woz00} Wozniak, P. 2000, Acta Astronomica, 50, 421.

\reference{Woz01} Wozniak, P., Udalski, A., Szymanski, M., Kubiak, M., Pietrzynski, G., Soszynski, I. and Zebrun, K. 2001, Acta Astronomica, 51, 175.

\end{references}
\end{document}